# Ontology Guided Information Extraction from Unstructured Text

Raghu Anantharangachar[1,] Srinivasan Ramani, S Rajagopalan

International Institute of Information Technology, Electronics City, Hosur Road, Bangalore 560100, India

`Raghu.a@iiitb.ac.in, Ramani.Srini@gmail.com, raj@iiitb.ac.in`

## ABSTRACT

*In this paper, we describe an approach to populate an existing ontology with instance information present in the natural language text provided as input. An ontology is defined as an explicit conceptualization of a shared domain [18]. This approach starts with a list of relevant domain ontologies created by human experts, and techniques for identifying the most appropriate ontology to be extended with information from a given text. Then we demonstrate heuristics to extract information from the unstructured text and for adding it as structured information to the selected ontology. This identification of the relevant ontology is critical, as it is used in identifying relevant information in the text. We extract information in the form of semantic triples from the text, guided by the concepts in the ontology. We then convert the extracted information about the semantic class instances into Resource Description Framework (RDF[3]) and append it to the existing domain ontology. This enables us to perform more precise semantic queries over the semantic triple store thus created. We have achieved 95% accuracy of information extraction in our implementation.*

## Categories and Subject Descriptors

I.2.7 [Natural Language Processing]: Text analysis

## General Terms

Algorithms

## 1. INTRODUCTION

Ontology based Information Extraction is a discipline in which the process of extracting information from various information repositories is guided by an ontology. The process of extraction of information itself involves multiple steps which include pre-processing the text into a machine processable form, and defining heuristics to identify the information to be extracted. The ability to extract information from text enables different applications such as question answering systems which can offer more precise answers – for example, a query like "provide me the list of all the papers written by X and Y in which Z is not an author" cannot be easily performed using existing information extraction techniques. The use of semantic information

---

existing in the sentences enables queries like this to be answered by search engines that use such information encoded in a suitable form. To cite another example, consider a user looking for information about a hotel in a certain locality. Semantic information can help interpret "being in a locality" in an appropriate way, using coded location information, reported distances from places known to be in a given locality, etc.

Ontology typically consists of two kinds of information items – those that make up A-Box and those that make up the T-Box. The T-Box consists of the terminology component which includes the definition of the classes, attributes, and their inter-relationships. The A-Box consists of assertions that make up the facts stated by the ontological instance. The A-Box basically consists of triples that provide information about various relationships that exist in an instance of the ontology, and the subject and object associated with each of those relationships (predicates). Our paper provides an approach and an implementation to extract all the A-box entries for the ontology given T-Box data for that particular ontology.

We focus on hotel domain here. We model information about hotels in our ontology, and extract information from the text to populate information about individual hotels
.

## 2. MOTIVATION

Natural language is the vehicle that carries the bulk of information used by us. Information in the text needs to be extracted from the text and converted to machine processable form in order to enable software applications to use this information for various purposes. The most significant question answering systems dealing with questions over textual data are modern search engines. Their strengths and weaknesses are widely appreciated. They are very good in locating documents and sentences in documents relevant to a given set of query words.

A relatively recent development in this field is the advent of semantic question answering systems which use formal semantic representation of information from natural language text to provide answers to various questions.

Sometimes the text being processed may not be grammatically correct, but it might still contain some information. For example, when we see "distance from airport: 17kms", we are able to recognize that this means that the distance from airport is 17 kms. This is not a strictly grammatical sentence in English, but it still contains useful information that readers find useful.

When a user needs to extract information from various data sources which spans over thousands of documents, it is difficult for humans to perform this extraction manually, and we need automated systems that can extract the information and make it available to the user. The advent of semantic web has made possible the creation of triples of information that can be queried using semantic web query languages like Simple Protocol and RDF Query Language (SPARQL)2. This has resulted in enabling the user to perform complex logical queries over the triple repository. A triple is a basic entity of semantic web that includes a subject, a predicate and an object.

Ontology based Information Extraction (OBIE) is a recent field that promotes techniques that use an ontology as an integral part of the system for extracting and presenting information from various data sources. Combining text processing, information retrieval and semantic web techniques we are able to extract useful knowledge from various text sources with reasonable accuracy.

## 3. ONTO-TEXT

## 3.1 System Architecture and Workflow for ontology instance extraction

The architecture diagram for our ontology based information extraction system is given below:

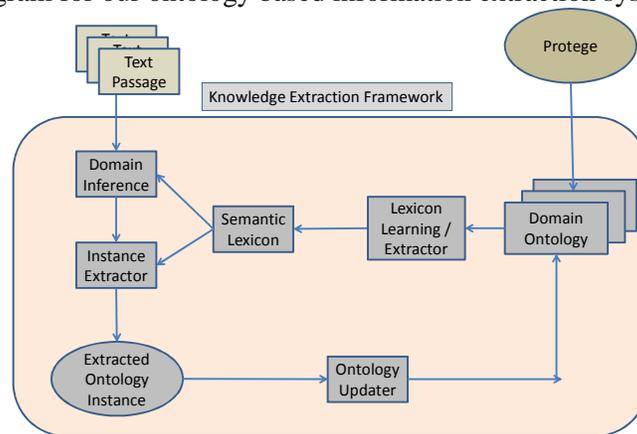

Figure 1: Knowledge Extraction Framework

This system uses a knowledge extraction procedure that reads a set of text documents, and extracts the ontology instances from those documents.

We use the concept of Semantic Lexicon to identify a semantic domain for the text being processed. A semantic lexicon is basically a set of words that are domain specific – they are an integral part of the domain vocabulary. A lexicon is a set of words and it usually is not specific to any domain. A Semantic Lexicon consists of words that identify a domain uniquely. For example, a semantic lexicon for a banking domain includes words such as account, savings account, current account, payee, transaction and the like. A Semantic Lexicon for a hotel domain includes words such as serve, meal, dine, buffet, dinner and the like.

Our approach includes reading input from a given text document, and then using a domain inference module that incorporates a semantic lexicon, to identify the domain. The semantic lexicon is predefined by experts for each domain based on their expertise in that particular domain.

Once the domain is identified, the instance extractor module extracts the instance information, and creates an RDF node, and updates the ontology. We use Jena APIs for this purpose. This ontology can be edited using any ontology editor such as Protégé[19]. The Lexicon learning/extractor module has rules to learn new lexicon symbols from the text, and add them into the semantic lexicon. The lexicon learning module uses a set of heuristics to identify lexical items that are related to the existing semantic lexicon.

We outline our pipeline for the extraction of ontology instances in the workflow below:

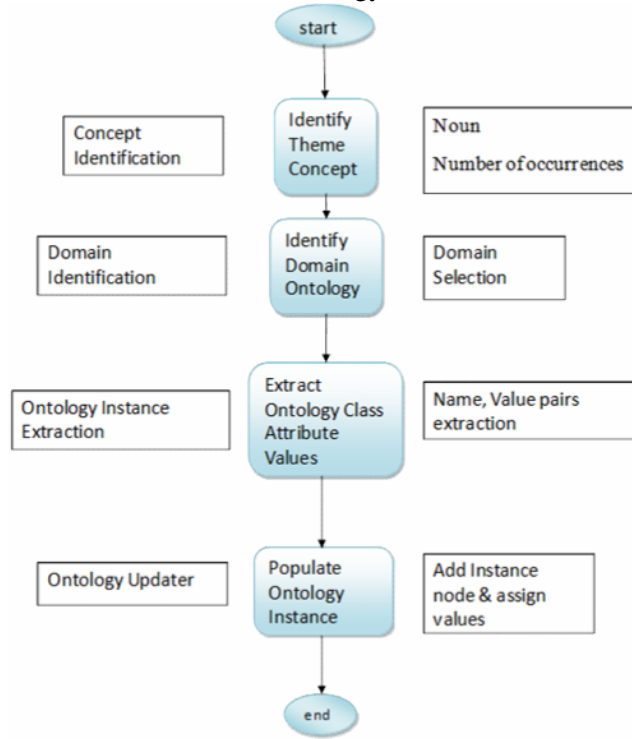

Figure 2: Knowledge Extraction Workflow

## Relevant Research Problems

Extracting ontology instance information from text is a difficult task, and it involves multiple problems being solved. Some of the problems are:

1. How do we identify the string representing the subject of a sentence?
2. How do we disambiguate the string and assign it to the appropriate semantic class? (Semantic class assignment problem). For example, if we find that Oberoi is the string that represents the subject in a text passage, then how do we identify the class to which it belongs (for example, hotel)?
3. How do we extract the values for the various attributes from the text?

## 3.2  Triplet Extraction Algorithm and Code

We extract the subject, predicate and object constituting a connected component of a sentence, and extract an assertion from the sentence that can be stored in the semantic store. We also identify attribute-name value pairs when we encounter known attributes. This approach uses Natural Language Processing (NLP) to identify a subject, and maps this subject to a semantic class, and uses the predicate and object as the attribute name and value respectively.

For example, consider a sentence like "Oberoi is located in Bangalore". In this sentence, the subject is Oberoi, the predicate is located, and the object is Bangalore. We use the subject as the name of the instance of the hotel class in our ontology, and use Bangalore as the value for the attribute named location. So, after parsing this sentence, we are able to fill up the attribute value form of location-information for the ontology instance of a hotel class called Oberoi. The following RDF snippet is added to the ontology after processing this sentence.

```
<Hotel rdf:ID="Oberoi">
    <hasLocation rdf:datatype="http://www.w3.org/2001/XMLSchema#string"
     >Bangalore</hasLocation>
</Hotel>
```

We use a triple extraction algorithm described in [Deli] that we have implemented using the StanfordCoreNLP[16] java library. The algorithm for extracting the subject, predicate and object is available in [15] and is briefly summarized below for convenience:

Algorithm ExtractSubject (string)
1. Perform a Breadth First Search (BFS) of the parse tree obtained by using StanfordCoreNLP library.
2. The NP subtree contains the subject, and it is the first Noun in the tree when traversed using BFS.

Algorithm ExtractPredicate (string)
1. Perform a Depth First Search of the VP subtree. The verb that is deepest in the tree is the predicate.

Algorithm ExtractObject (string)
1. Perform a search of the PP, ADJP subtree, and extract the first noun in the tree. This is the object.

We have implemented an initial version of the above algorithm using java and StanfordNLP package. We find our implementation to work for simple sentences (sentences that have a single subject, single predicate and single object), and needs to be enhanced for more complex sentences.

## 3.3  Hotel Ontology

Our approach is to examine a sizable set of documents and evolve a process to handle the various types of text pages available about each hotel, for extracting information from these pages accurately. We have evolved an ontology for hotels to model the important attributes and classes that we think are appropriate for people looking for information from these pages. The various attributes of hotel class are listed here.

```
Hotel
    shoppingArcade
    fitnessCenter
    Jacuzzi
    petsAllowed
    poolsideBar
    valetParking
    carPark
    internetAccess
    tours
    smokingArea
    shuttleService
    salon
    nightclub
    meeting_facility
    dry_cleaning
    laundry
    family_room
    executive_floor
    disabled_facilities
    concierge
    coffee_shop
    business_center
    bicycle_rental
    bar_pub
    babysitting
    airport_transfer
    garden
    steamroom
    safedepositbox
    elevator
    roomservice
    numroomservice
    restaurant
    indoorpool
    outdoorpool
    swimmingpool
    pool
    gym
    casino
    fitnessCentre
    spa
    nightlifelounge
    rooms
    numrooms
    acres
    numacres
    airport
    distfromairport
    distfromrailstn
    distfrombusstand
    star
    numstars
    tariff
    tariffamt
    contact
    contactstr
    facility
    facilitystr
    ambience
    ambiencestr
```

**Figure 3: Hotel Ontology**

## 3.4  Our Approach

As there is no perfect way of extracting the entire information from the text without any ambiguity, we use an approach that increases our chances of extracting the most information from the text correctly. In order to do this, we use multiple heuristics.

The first step in the entire process is the identification of the most important string in the text passage. This string forms the key string which basically refers to the central topic of the entire passage. We call this string as the theme concept. For example, if a paragraph of text is talking about Oberoi, then Oberoi is the theme concept. The following paragraph describes how to identify the theme concept for a given text paragraph.

### 3.5.1 Theme concept identification

We believe that identifying the theme concept can increase our chances of interpreting the text in the paragraph . For example, if the theme concept is a hotel, then the text in the paragraph is interpreted as giving values for the attributes of a hotel instance. Similarly, if the theme concept is a hospital, then the text in the paragraph is interpreted as giving values for the attributes of a hospital. The theme concept is identified as follows.

Step 1:
We define Concept as the set of all nouns in the paragraph which is populated in the following manner:

We perform sentence parsing, tokenization and parts-of-speech tagging on the given input text using Stanford NLP package. Then, we extract all the tokens that are tagged as nouns in the text and populate the Concept set. Each member of this set is called as a concept.

Mathematically, Concept = { nouns }

Step 2: Extract the triplets from the sentence, and construct the subejectList from the list of subjects extracted.
subjectList = {subjects}

Step 3: Compute the number of occurrences of each concept in the sentence. Identify the concepts that occur the maximum number of times, from this list. This set of concepts that occur the maximum number of times in the given text passage, form the MaxOccurConcepts set.

Mathematically we can represent the MaxOccurConcepts set as follows:
MaxOccurConcepts = {concept}  Each concept is a noun that occurs in the text passage.

Step 4: Identify the Theme concept. We define the Theme concept as the concept that is described in the text. It can be identified as being the subject in one or more sentences in the text fragment, and which occurs the largest number of times in the text.
Mathematically, we can represent the Theme concept as:
Themeconcept = Concept     subjectList    MaxOccurConcepts

Then, whichever concept satisfies all the above mentioned conditions, gets flagged as theme concept.

Consider the following text passage:Shantiniketan at Prince Street offers excellent accommodation for the guests. Comprising of 6 blocks and 56 deluxe rooms, Shantiniketan offers a decent stay along with delectable delights of the Princess Café.

Concepts = {Shantiniketan, Prince Street, accommodation, guets, blocks, rooms, stay, Princess, Café}
subjectList = {Shantiniketan}
MaxOccurConcepts = {Shantiniketan (2)}

ThemeConcept = {Shantiniketan}

We identify Shantiniketan as the theme concept and discover other newer concepts which are related to this concept, and as well extract the values for the various attributes from the text.

### 3.5.2 Domain Ontology Identification

Having identified the theme concept, the next step for us is to understand the domain to which the theme concept belongs. This step is required to identify an appropriate domain ontology that we can use. We believe that when we are extracting information from text there could be multiple domains involved – for example, we could be extracting information about hotels, or we could be extracting information about hospitals and so on. We need to know which domain ontology to use ( for example, whether to use hotel ontology or to use hospital ontology) while extracting information from text.

We have two rules to identify the domain class to which the string identified as Theme concept corresponds to. These are:

Explicit mention Rule: The occurrence of the strings that are the names of the class itself – for example, if a string hotel appears in the text, it is likely that the string is talking about a hotel class.

Implicit lexicon match Rule: If there are no explicit mentions of the class names in the string, then we use this approach. We use a domain ontology lexicon which is derived from the ontology class attribute names in the ontology, and perform a string matching. We assume the existence of a domain ontology that is created by experts in that particular domain. This is the usual practice that we have seen, in that ontologies are usually hand-crafted in enterprises. Each ontology has a set of classes, subclasses, attributes and relationships. By using this ontology as the starting point, we create a semantic lexicon that is relevant for each domain. We use this semantic lexicon to identify the domain to which the text passage refers. We also assign weights to each of these names, attributes and relations so that we are able to infer the domain to which the passage refers with a certain degree of confidence. We define the semantic lexicon as a set containing all these attribute names, class names and their relation names with associated weights.

Then, we find the number of matches of the semantic lexicon for each domain with the text passage. Based on the number of matches, we decide the domain to which the text passage refers to.

Additionally, when people are looking for specific local information using a yellow page service provider, the users usually provide the details about the entity of interest which is being queried. In our earlier work [14], we have created a prototype yellow page service system that uses a form based input to disambiguate the user's intent. This tells us about the domain to which the query belongs, which is also the ontology we are interested in.

Once we have the information about the above mentioned two sets of information, we take an intersection of these sets, and this tells us about the domain ontology to be used for ontology instance extraction. We then choose the domain ontology and load it into memory.

For example, in the example given earlier, there is an explicit use of the word "hotel". Hence the string Shantiniketan is likely to be referring to a hotel instance.

We create the hotel domain lexicon by using the names of the various classes/attributes/relations that are present in the ontology. For example, a sample hotel domain lexicon is given below:
Hotel domain Lexicon = {serve, food, breakfast, lunch, dinner, buffet}
Similarly, a simplified hospital domain lexicon is:
Hospital domain lexicon = {surgery, operation, doctor, patient}
We find no matches against the hospital domain lexicon, whereas the with the hotel domain lexicon, the number of matches is 2..

So, in our case, the string Shantiniketan refers to a Hotel instance.

### 3.4.3 Extraction of Ontology Attribute Values

We now extract the values of the various attributes that are part of the hotel class. In the earlier example, these include name, location, cost and serve.

We have,

Name = Shantiniketan
Facilities = accomodation
rooms = 56

Once these values are filled into the hotel instance node, we include the instance into the ontology. At this time, we have a new node representing a hotel instance class with all the attribute values present in the text.

3.5.1 Attribute Value extraction methodology
We use pattern matching technique to extract the name value pairs from text. A pattern contains a few terms and a set of constraints on those terms and when the pattern is matched, the rule is executed. We classify our patterns into three categories:

3.5.1.1 Simple
Simple patterns are very easy to identify and extract. They only match an occurrence of a string. For example, occurrence of a string "doctor-on-call" implies that a doctor-on-call is present in the hotel.

3.5.1.2 Medium
A medium pattern includes a string and a number that occurs either before or after the string. For example, while extracting number of rooms a pattern of the type of "[n] rooms" is useful to match.

3.5.1.4 Complex
A complex pattern includes any pattern that has multiple constraints in terms of starting string, ending string, and some intermediate strings as well. An example is when we are extracting the distance of a hotel from airport, we could specify a pattern that has the following form:
[n] kms from airport

Or it could also be:
Distance from [text] airport is [n]

In this pattern, there could be multiple text terms that might appear between the "from" and "airport". For example, the string could read as "distance from BIAL airport is 10kms" or could read as "distance from the nearest airport is 10kms". Our pattern matching software extracts all patterns that match this pattern, and return the name, value pairs. For example, in the above example, it returns "name=distance from airport", "value=10kms".

### 3.4.4 Ontology Update

We convert the extracted values to RDF format, and append it to the hotel ontology. We use the open-source Jena tool (http://jena.apache.org/) from HPLabs for updating the ontology. We initially load the ontology using Jena APIs into memory. And, then, we create a new ontology instance node and assign the various values that were extracted from the text. Then, we invoke jena APIs to update the ontology.

The RDF representation of the extracted information given in the earlier paragraph is given below:
<Hotel rdf:ID="Shantiniketan">
 <Facilities rdf:datatype="http://www.w3.org/2001/XMLSchema#string"
  >accomodation</location>
<rooms rdf:datatype="http://www.w3.org/2001/XMLSchema#integer"
  >56</serves>
</Hotel>

### 3.5 Querying using SPARQL queries

Once we have input all the text about the various classes defined in our ontology, our system creates a semantic store which is basically an updated ontology with which we started. The initial ontology is created by human experts, and it is made to represent an actual real life requirement for a particular domain like hotel. In our case, we started with a hotel ontology, and then ran the various text documents thru our system, and ended up with an updated hotel ontology. Then, we are able to run SPARQL queries by loading this ontology in Protégé. We are able query for various triples, and retrieve the results using SPARQL. Our implementation demonstrates that our ontology instance population algorithm can perform extraction of ontology instance information from the given natural language text, and append it to an existing ontology. We also have a form based tool using which we can query the information in the semantic store which is already published as part of another paper[14].

## 4. Related Work

The various fields/communities that are related to our work described in this paper are natural language processing, linguistic processing, artificial intelligence, semantic web, java communities and open source communities.
Natural Language Processing / Linguistic Processing:

Researchers in the NLP field have researched for the past several decades in the field of extracting information from natural language text [10]. They have been adopting various approaches mostly centered on syntactic parsing of sentences and pattern matching using rules. The syntactic approaches are of limited use as they look for grammar in the sentences, and use syntactic clues to extract information from text. Pattern matching using Hearst rules [11] also mostly works on similar principles, but provides an accurate extraction of information when the rules can be enumerated. The linguistic processing community has worked on techniques for understanding linguistic structures of sentences, and applying these techniques for information extraction. However, these approaches run into problems when the linguistic patterns conflict or lack completeness. The linguistic approaches also use the sentence parse trees to understand the various components of the sentences, understand their parts of speech, and then use this understanding to extract information from text. When applied individually, the approach is not able to extract the entire information from sentences. [7] is about a protégé plug-in which uses linguistic analysis to extract ontological instance information. [8] extracts relations from large collections of plain-text documents. [12] is an attempt to extract structured and semi-structured data from natural language text. [13] is an example for learning rules for information extraction from unstructured text.

Artificial Intelligence and Databases:
The Artificial Intelligence community has looked at understanding of given natural language text and have evolved natural language understanding systems which when trained well provide limited understanding of the sentences, and based on their reasoning abilities, are able to extract information from sentences. There are efforts from the database community to create systems that can populate the schema elements given a pre-defined schema in a relational database. There are also efforts to extract relationship information from various schema elements. These systems assume a pre-existing schema in most cases, and when they try to learn the schema itself, the results are not so encouraging.

Semantic web:
The semantic web community has looked at defining and learning ontologies for information extraction and this approach has enabled extraction of reasonably accurate information from text, given a good ontology. There are approaches that attempt to learn an ontology from natural language text and these systems are still in their infancy: they try to build an ontology from given text, which can be used for purposes of information extraction. Using an ontology for guiding the information extraction process itself is a new and emerging field called as Ontology Based Information Extraction (OBIE), and offers the promise of providing good accuracy of information extraction. Researchers have used an ontology for both guiding the information extraction itself, and for presenting the results of the extraction as well.

[1] is an early effort in using ontologies for information extraction and uses constants and keywords to extract information, which is then generalized. [2] presents a survey of the OBIE field. SOBA[3] is another effort towards extracting information from soccer pages for question answering. KIM[4] uses an upper level ontology for information extraction. [5] proposes metrics for evaluation of extracted ontologies. [6] provides an application of ontology based information extraction for business intelligence. TextOntoEx[9] deals with creating ontology from natural language text, using a semantic pattern based approach.

Open source and Java community:
The Open Source community has evolved tools that can be used for information extraction, such as Apache Lucene ( http://lucene.apache.org/core/) that supports proximity queries for extracting information from text, in addition to providing a free text search capability. Our methodology is generic and can be applied to any domain and is based on Ontology based Information Extraction. We use ontology to guide us in the information extraction process, and we also use semantic processing to identify the domain. Hence, our approach is hybrid, and is novel.

## 5. Evaluation and Metrics

We observe that the Learning Accuracy metric defined in [20] is too broad and is more relevant for ontology learning scenarios. So, we define new metrics named Ontology Coverage and Ontology Attribute Accuracy.

Ontology Coverage provides an indication of how many of the fields that are present in the text have been successfully extracted. It is a measure of the precision of the extraction process itself. Ontology Coverage = total number of attributes extracted / total number of attributes present in the text

Ontology Attribute Accuracy indicates how correctly the various attributes have been extracted. For example, if there is a location field present in the text, but it is not extracted correctly, then it reduces the value of this metric.

Basically, the coverage is a measure of how many of the attributes have been extracted, and the accuracy is a measure of how correctly the various attributes have been extracted.

Ontology Attribute Accuracy = number attributes extracted correctly / total number of attributes extracted from the text Semantic Domain Accuracy is a measure of how correctly the string encountered in the text is mapped to an appropriate semantic domain. For example, if Oberoi is encountered in the text, and it is mapped to hotel domain, then the accuracy is 100%. Otherwise it is 0%. The reason why this accuracy is binary in nature is because, if we use the wrong ontology, then the extraction is not going to give us good results.

Theme Concept Accuracy is a measure of if the correct string is identified for the theme concept or not. This is also a binary value and is 100% if the theme concept is correctly identified and 0% otherwise.

## 6. EXPERIMENTAL DESIGN

An experimental procedure was designed to validate our approach of using rules to carry out pattern matching to extract information from a large number of text pages. Our starting hypothesis was – if we use the ruleset that is designed manually to extract the information from a set of adequate size of text pages, it is possible to achieve an accuracy of more than 90% with any number of pages. We believed that adequate size would be of the order of 50 pages. We expected that it would be possible to process large number of text pages with this approach, and still get more than 90% accuracy. These text pages usually correspond to web pages, and it is possible to covert the web pages to text format, and process them to extract various ontology fields from the page.

## 7. EXPERIMENTS AND RESULTS

We have downloaded information from available public web sites of various hotels, and have applied our pattern matching technique to extract the information from those web sites. The results of our experiments are presented in the next page in the form of a table. The table contains the name of the site, the number of names present, and the number of the attributes correctly extracted from the site. As we can see, the approach has worked very well, and has resulted in very high accuracy.

We used the first round of inputs to train the extractor, and to refine the rules for extraction. We initially downloaded 52 text pages from web describing various hotels from various sources like the hotel web pages, review pages, blogs and so on. Then, we manually created rules for extraction of various attributes in the ontology. Once we did this, we ran the extractor on this set of pages. We fine-tuned the rules/heuristics so that extraction accuracy is improved. The results obtained in this set are given in Table 1. The columns in the table depict the total number of attributes retrieved, the number of attributes extracted correctly, and the number of attributes extracted incorrectly. The extraction accuracy is the percentage of number of correct extractions. The average accuracy during this round is found to be 99.82%. This gives us a confidence that we can now use the same set of rules against various other text pages.

| Sl.No | Hotel Actual name | Total Retrieved | Correctly Retrieved | Incorrectly Retrieved | Extraction Accuracy |
|---|---|---|---|---|---|
| 1 | arra Suites | 17 | 17 | 0 | 100.00 |
| 2 | Carlson | 3 | 3 | 0 | 100.00 |
| 3 | casa cavole | 8 | 8 | 0 | 100.00 |
| 4 | clarks exotica | 24 | 24 | 0 | 100.00 |
| 5 | comfort inn | 7 | 7 | 0 | 100.00 |
| 6 | convergence | 9 | 9 | 0 | 100.00 |
| 7 | empire | 6 | 6 | 0 | 100.00 |
| 8 | fantasy golf resort | 1 | 1 | 0 | 100.00 |
| 9 | geo | 12 | 12 | 0 | 100.00 |
| 10 | golsfinch | 28 | 28 | 0 | 100.00 |
| 11 | hadoti | 11 | 11 | 0 | 100.00 |
| 12 | hopperstop | 9 | 9 | 0 | 100.00 |
| 13 | atria | 13 | 13 | 0 | 100.00 |
| 14 | mayflower | 11 | 11 | 0 | 100.00 |
| 15 | india builders | 1 | 1 | 0 | 100.00 |
| 16 | ista | 17 | 17 | 0 | 100.00 |
| 17 | windsor sheraton | 9 | 9 | 0 | 100.00 |
| 18 | kabini serai | 10 | 10 | 0 | 100.00 |
| 19 | kanha earth lodge | 11 | 10 | 1 | 90.91 |
| 20 | komfort terraces | 6 | 6 | 0 | 100.00 |
| 21 | lalit resort | 12 | 12 | 0 | 100.00 |
| 22 | lalith mahal | 3 | 3 | 0 | 100.00 |
| 23 | le merridien | 15 | 15 | 0 | 100.00 |
| 24 | luciya intl | 11 | 11 | 0 | 100.00 |
| 25 | museum inn | 7 | 7 | 0 | 100.00 |
| 26 | nahar heritage | 8 | 8 | 0 | 100.00 |
| 27 | nalapad | 11 | 11 | 0 | 100.00 |
| 28 | nandhini | 9 | 9 | 0 | 100.00 |
| 29 | neemrana | 10 | 10 | 0 | 100.00 |
| 30 | oberoi amarvilas | 10 | 10 | 0 | 100.00 |
| 31 | ramada | 7 | 7 | 0 | 100.00 |
| 32 | ramanashree | 1 | 1 | 0 | 100.00 |
| 33 | ramanashree blr | 11 | 11 | 0 | 100.00 |
| 34 | ramee guestline | 12 | 12 | 0 | 100.00 |
| 35 | royal orchid central | 8 | 8 | 0 | 100.00 |
| 36 | royal orchid harsha | 1 | 1 | 0 | 100.00 |
| 37 | sai leela | 28 | 28 | 0 | 100.00 |
| 38 | savannah sarovar | 4 | 4 | 0 | 100.00 |
| 39 | solitaire | 1 | 1 | 0 | 100.00 |
| 40 | st marks | 6 | 6 | 0 | 100.00 |
| 41 | taj residency | 13 | 13 | 0 | 100.00 |
| 42 | taj vivanta | 10 | 10 | 0 | 100.00 |
| 43 | taj westend | 3 | 3 | 0 | 100.00 |
| 44 | capitol | 10 | 10 | 0 | 100.00 |
| 45 | gateway | 12 | 12 | 0 | 100.00 |
| 46 | lali ashok | 5 | 5 | 0 | 100.00 |
| 47 | leela palace | 3 | 3 | 0 | 100.00 |
| 48 | oberoi group | 5 | 5 | 0 | 100.00 |
| 49 | the park | 1 | 1 | 0 | 100.00 |
| 50 | the pride | 3 | 3 | 0 | 100.00 |
| 51 | vivanta | 4 | 4 | 0 | 100.00 |
| 52 | wildflower | 11 | 11 | 0 | 100.00 |

Table 1: Training Data

Accuracy (Average) = 99.82%

In the second step, we applied the same set of rules and the extraction techniques against another 40 different text pages again downloaded from various hotel sites, review sites, blogs, etc to ensure that we have diversity in our input documents. The results obtained are given in Table 2.

| Sl.No | Hotel Actual Name | Total Retrieved | Correctly Retrieved | Incorrectly Retrieved | Extraction Accuracy |
|---|---|---|---|---|---|
| 1 | Vythri resorts | 9 | 9 | 0 | 100.00 |
| 2 | Annamali intl | 7 | 7 | 0 | 100.00 |
| 3 | Atithi | 9 | 9 | 0 | 100.00 |
| 4 | fortune kences | 6 | 6 | 0 | 100.00 |
| 5 | Bliss | 8 | 8 | 0 | 100.00 |
| 6 | sindhuri park | 2 | 2 | 0 | 100.00 |
| 7 | Kalyan residency | 8 | 8 | 0 | 100.00 |
| 8 | grand | 9 | 9 | 0 | 100.00 |
| 9 | lakeview | 5 | 5 | 0 | 100.00 |
| 10 | westwood munnar | 10 | 9 | 1 | 90.00 |
| 11 | igloo munnar | 6 | 6 | 0 | 100.00 |
| 12 | vythri windflower | 9 | 9 | 0 | 100.00 |
| 13 | atria | 7 | 7 | 0 | 100.00 |
| 14 | best westernfoot | 9 | 8 | 1 | 88.89 |
| 15 | t&u munnar | 10 | 9 | 1 | 90.00 |
| 16 | resort tea county | 8 | 8 | 0 | 100.00 |
| 17 | cloud9 resorts | 8 | 7 | 1 | 87.50 |
| 18 | abad copper castle | 8 | 8 | 0 | 100.00 |
| 19 | chancellor resorts | 8 | 8 | 0 | 100.00 |
| 20 | sterling resorts | 9 | 8 | 1 | 88.89 |
| 21 | lakeview resort | 7 | 7 | 0 | 100.00 |
| 22 | paradise plantation retreat | 7 | 5 | 2 | 71.43 |
| 23 | windermere estate | 9 | 8 | 1 | 88.89 |
| 24 | oakfields | 12 | 11 | 1 | 91.67 |
| 25 | naini retreat | 7 | 7 | 0 | 100.00 |
| 26 | new bharat | 5 | 5 | 0 | 100.00 |
| 27 | himalaya hotel | 7 | 7 | 0 | 100.00 |
| 28 | elphinestone hotel | 3 | 3 | 0 | 100.00 |
| 29 | fairhavens hotel | 5 | 5 | 0 | 100.00 |
| 30 | manu maharani | 10 | 10 | 0 | 100.00 |
| 31 | sherwani hilltop | 9 | 9 | 0 | 100.00 |
| 32 | dynasty resort | 4 | 4 | 0 | 100.00 |
| 33 | alka nainital | 4 | 4 | 0 | 100.00 |
| 34 | green magic nature | 2 | 2 | 0 | 100.00 |
| 35 | vikram vintage | 6 | 6 | 0 | 100.00 |
| 36 | ginger pondicherry | 11 | 11 | 0 | 100.00 |
| 37 | lotus warm | 7 | 7 | 0 | 100.00 |
| 38 | sarovar hotels | 1 | 1 | 0 | 100.00 |
| 39 | du parc | 4 | 3 | 1 | 75.00 |
| 40 | richmond hotel | 4 | 4 | 0 | 100.00 |

Table 2: Data used in first round of Extraction

Accuracy (Average) = 96.81%

As we can see from this table, the extraction accuracy is 96.81%. This implies that without changing the rules, we are able to apply our technique to a fresh set of text pages, and get a very good extraction accuracy.

Having obtained excellent results, we decided to validate our rules and extraction technique against another set of 40 text pages. So, we downloaded again another set of pages, and re-ran the same extractor against these pages. The results are given in Table 3. As we can see, the extraction accuracy is 95.23%. This implies that our initial set of rules along with the extraction technique is robust enough to be applied to a variety of text pages, and still gives an excellent extraction accuracy (over 95%).

| Sl.No | Hotel Name | Total Retrieved | Correctly Retrieved | Incorrectly Retrieved | Extraction Accuracy |
|---|---|---|---|---|---|
| 1 | beach symphony | 4 | 3 | 1 | 75.00 |
| 2 | Amritha castle | 14 | 13 | 1 | 92.86 |
| 3 | Breeze residency | 7 | 6 | 1 | 85.71 |
| 4 | cambay grand | 10 | 10 | 0 | 100.00 |
| 5 | courtyard ahmedabad | 11 | 11 | 0 | 100.00 |
| 6 | fortune landmark | 8 | 8 | 0 | 100.00 |
| 7 | fortune resort bay island | 2 | 2 | 0 | 100.00 |
| 8 | gemini continental | 2 | 2 | 0 | 100.00 |
| 9 | grand gardenia | 10 | 9 | 1 | 90.00 |
| 10 | clarks awadh | 2 | 2 | 0 | 100.00 |
| 11 | grand central bhubhaneswari | 8 | 8 | 0 | 100.00 |
| 12 | newyork hotel | 5 | 5 | 0 | 100.00 |
| 13 | sarovar portico | 9 | 9 | 0 | 100.00 |
| 14 | lemeridien ahmedabad | 9 | 9 | 0 | 100.00 |
| 15 | lemon tree hotel ahmedabad | 8 | 8 | 0 | 100.00 |
| 16 | mintokling | 8 | 8 | 0 | 100.00 |
| 17 | peerless resort | 3 | 3 | 0 | 100.00 |
| 18 | royal orchid central ahmedabad | 10 | 10 | 0 | 100.00 |
| 19 | sarovar portico | 16 | 16 | 0 | 100.00 |
| 20 | shantiniketan | 6 | 5 | 1 | 83.33 |
| 21 | sliversand beach resort | 2 | 2 | 0 | 100.00 |
| 22 | sitara luxury | 9 | 8 | 1 | 88.89 |
| 23 | st lauren towers | 5 | 5 | 0 | 100.00 |
| 24 | swosti premiums | 11 | 10 | 1 | 90.91 |
| 25 | tara comfort | 12 | 11 | 1 | 91.67 |
| 26 | chambers hotel | 5 | 4 | 1 | 80.00 |
| 27 | crown bhubhaneswar | 12 | 12 | 0 | 100.00 |
| 28 | fern ahmedabad | 14 | 14 | 0 | 100.00 |
| 29 | gateway hotel ahmedabad | 12 | 11 | 1 | 91.67 |
| 30 | house of mg | 7 | 7 | 0 | 100.00 |
| 31 | pride hotel | 9 | 8 | 1 | 88.89 |
| 32 | sikkim | 12 | 12 | 0 | 100.00 |
| 33 | srm hotel | 13 | 13 | 0 | 100.00 |
| 34 | westin hyderabad | 13 | 11 | 2 | 84.62 |
| 35 | trident bhubhaneswar | 10 | 10 | 0 | 100.00 |
| 36 | tulip inn ahmedabad | 5 | 4 | 1 | 80.00 |
| 37 | vasundhara villa | 7 | 6 | 1 | 85.71 |
| 38 | shivamurugan | 6 | 6 | 0 | 100.00 |
| 39 | singaar | 12 | 12 | 0 | 100.00 |
| 40 | sparsa resort | 7 | 7 | 0 | 100.00 |

Table 3: Data used in second round of Extraction

Accuracy (Average) = 95.23%

## 8. DISCUSSIONS

As we see from the tables above, the accuracy of extraction for the new pages is 95%, which is very good. The drop in the accuracy from 99% to 95% is because there are few additional usage patterns that are present in this text, which are not found in the training samples. However, by adding a few more rules (5 to be precise), we were able to increase the accuracy to 99%. This reconfirms our hypothesis that the number of rules and patterns that need to be created to extract the information correctly from various pages is finite and manageable in number, and is possible to hand-craft these patterns manually in a short period of time.

## 9. CONCLUSION AND FUTURE WORK

We believe and argue that our suggested methodology is quite generic, and can be easily adapted to any new domain (though we used the hotel domain to validate our approach). We believe that the methodology of defining rules for patterns is quite reasonable and simple and can be easily created by humans who are experts in that particular field. And, also there are only a limited number of rules that need to be created to cater to most of the cases. Hence, the effort for creating new rules is very much justified.

An Application of this type becomes valuable when used across a large number of web sites (say a few thousand of them) to extract relevant information and add it to an ontology. The updated ontology can then be used to handle semantic queries which otherwise would not be possible. As mentioned earlier, we have reported elsewhere [] techniques giving a simple form-based interface, which generates semantic queries automatically.

## 10.  ACKNOWLEDGMENTS

We would like to thank and acknowledge the support provided by HP Labs India. We also would like to thank and acknowledge the support provided by International Institute of Information Technology.